\documentclass{article}

\usepackage{PRIMEarxiv}
\usepackage{booktabs} 
\usepackage{array}
\usepackage{multirow}
\usepackage{enumitem}
\usepackage{colortbl}
\usepackage{wasysym}
\usepackage{xcolor}
\usepackage[normalem]{ulem}

\usepackage[utf8]{inputenc} 
\usepackage[T1]{fontenc}    
\usepackage{hyperref}       
\usepackage{url}            
\usepackage{booktabs}       
\usepackage{amsfonts}       
\usepackage{nicefrac}       
\usepackage{microtype}      
\usepackage{lipsum}
\usepackage{fancyhdr}       
\usepackage{graphicx}       
\graphicspath{{media/}}     

\usepackage{booktabs} 
\usepackage{array}
\usepackage{multirow}
\usepackage[export]{adjustbox}
\usepackage{pifont}
\usepackage{subfigure}
\usepackage{enumitem}

\usepackage[normalem]{ulem}
\usepackage{setspace}

\usepackage{geometry}
\usepackage{tikz}
\usetikzlibrary{arrows.meta,positioning}

\usepackage{booktabs}
\usepackage{tabularx}
\usepackage{array}
\usepackage{ragged2e}
\usepackage{xcolor}
\newcolumntype{Y}{>{\RaggedRight\arraybackslash}X}

\usepackage{multirow}

\title{Building AI Literacy at Home: How Families Navigate Children’s Self-Directed Learning with AI
}

\author{
  Jingyi Xie\textsuperscript{1}\thanks{Both authors contributed equally to this research.}, 
  Chuhao Wu\textsuperscript{2}\footnotemark[1],
  Ge Wang\textsuperscript{3}, 
  Rui Yu\textsuperscript{4},
  He Zhang\textsuperscript{5},  
  Ronald Metoyer\textsuperscript{6}, 
  Si Chen\textsuperscript{6}\thanks{Corresponding author: schen34@nd.edu} 
  \\
  {1} Department of Industrial and Systems Engineering, San José State University
  \\
  {2} University Libraries, Clemson University
  \\
  {3} Siebel School of Computing and Data Science, University of Illinois at Urbana-Champaign
  \\
  {4} Department of Computer Science and Engineering, J.B. Speed School of Engineering, University of Louisville
  \\
  {5} College of Information Sciences and Technology, Pennsylvania State University
  \\
  {6} University of Notre Dame
}

\begin{document}
\maketitle

\begin{abstract}

As generative AI becomes embedded in children’s learning spaces, families face new challenges in guiding its use. Middle childhood (ages 7-13) is a critical stage where children seek autonomy even as parental influence remains strong. Using self-directed learning (SDL) as a lens, we examine how parents perceive and support children’s developing AI literacy through focus groups with 13 parent–child pairs. Parents described evolving phases of engagement driven by screen time, self-motivation, and growing knowledge. While many framed AI primarily as a study tool, few considered its non-educational roles or risks, such as privacy and infrastructural embedding. Parents also noted gaps in their own AI understanding, often turning to joint exploration and engagement as a form of co-learning. Our findings reveal how families co-construct children’s AI literacy, exposing tensions between practical expectations and critical literacies, and provide design implications that foster SDL while balancing autonomy and oversight.


\end{abstract}


\section{Introduction}

Self-directed learning (SDL) has long been seen as a cornerstone of autonomy and lifelong learning \cite{knowles1975self,tough1979adult,zimmerman2002self}. For children, SDL is especially important as they move through critical developmental stages of building independence, metacognitive awareness, and executive functions such as planning, attention, and reflection \cite{winne2011self,paris2001metacognition,schunk2014handbook}. At this age, learners are beginning to take responsibility for identifying goals and monitoring their progress, yet remain particularly vulnerable to over-reliance on external aids. Generative AI has recently emerged as a powerful support for SDL, promising to provide instant explanations, adaptive feedback, and personalized resources that extend learning beyond school curricula \cite{kim2025impact,ali2023teachergaia,ouaazki2024generative}. Indeed, recent surveys show that ChatGPT has rapidly become one of the most widely used tools for learning globally, ranking just below YouTube in popularity \cite{toptools_survey}. In principle, generative AI could open new opportunities for children to explore knowledge, practice reflection, and pursue interests that traditional schooling cannot fully support.

At the same time, the affordances of generative AI come with substantial risks. Because these systems are designed to produce fluent, persuasive natural language, children may be especially prone to over-reliance on their outputs, short-cutting critical reflection and problem-solving \cite{dizon2024chatgpt,wang2024understanding,li2025impact}. Hallucinations and misinformation present new dangers, as does the tendency of generative AI to mimic empathy and conversational depth that may blur boundaries between tool and companion \cite{yu2025exploring}. Recent reports of harmful child-AI interactions, including tragic cases linked to suicide, underscore the ethical imperative to design systems that are safe, developmentally appropriate, and socially responsible \cite{parents_chatgpt_suicide}. These concerns have spurred mounting pressure: OpenAI has committed to introducing parental controls (e.g., account linking, real-time distress alerts, and age-appropriate response rules) within weeks as part of a broader safety initiative \cite{openai_parental_controls}. Meanwhile, U.S. regulators are investigating the impact of generative AI on children’s mental health, signaling a broader call for oversight \cite{ftc_investigation}.

Schools, however, often lag behind technological adoption and regulation, leaving parents as the first line of mediation \cite{druga2022family,andries2023alexa}. Prior work on children’s AI use has highlighted parents’ roles, but often in relation to earlier technologies. For example, Druga et al. examined families with children ages 5--12 to show how parents and children co-construct AI literacy in home settings using smart toys and speakers \cite{druga2022family}. Andries et al. studied children ages 6--10 with smart speakers, illustrating misconceptions of AI’s abilities \cite{andries2023alexa}, while Sun et al. investigated parents of preschoolers (ages 3--5) mediating early AI storytelling experiences \cite{sun2024parents}. In contrast, most SDL research on generative AI has targeted undergraduates or adolescents \cite{biyiri2024chatgpt,esiyok2025acceptance}, with only a handful exploring children’s direct use of generative AI for language learning or computational thinking \cite{dizon2024chatgpt,ouaazki2024generative}. This leaves little understanding of the middle childhood and early adolescence period (ages 7--13, corresponding roughly to grades 2--8 in the U.S. system), where parental influence remains strong but children are beginning to exercise greater agency in their technology use. In this paper, we investigate how parents of children ages 7--13 perceive and mediate generative AI use for learning. We ask:  
\textbf{RQ1.} How do parents describe phased pathways of children’s use of AI?  
\textbf{RQ2.} What strategies do parents adopt to mediate children’s use of AI for SDL?  
\textbf{RQ3.} How do parents conceptualize their role in supporting children’s SDL with AI?

This paper contributes to HCI in three ways. First, we articulate how parents conceptualize children’s AI literacy and its progression during ages 7--13 in the context of the pervasive presence of generative AI, extending existing frameworks that have primarily focused on either younger children (ages 3--6) or older learners (high school and beyond) \cite{long2020what,ng2021conceptualizing}, and developed at a time when AI was less prevalent. Second, we describe the concrete strategies parents adopt to monitor and guide children’s interactions with generative AI in SDL. Third, we provide design implications for child-development-friendly AI systems that move beyond age-based restrictions, highlighting adaptive and formative approaches aligned with parental expectations. Together, these contributions advance responsible AI design that supports children’s SDL while safeguarding their development.

\section{Background and Related Work}

\subsection{Children's AI Literacy}
AI literacy is a multifaceted and evolving concept. The literature generally converges on four central dimensions: a foundational understanding of AI systems, the effective usage and application of AI, the ability to critically evaluate the technologies and outputs, and attention to ethical considerations \cite{ng2021conceptualizing}. For instance, Long and Magerko~\cite{long2020ai} describe AI literacy as a set of competencies that enable individuals to understand AI concepts, recognize the limitations of such systems, and assess their broader impact. Pinski and Benlian~\cite{pinski2024ai} conclude that AI literacy is conceptualized as a holistic proficiency concept in 5 dimensions: knowledge, awareness, skills, competencies, and experience, which collectively contribute to the enabling and human-focused literacy
construct. Despite commonalities, scholars increasingly recognize that the requirements for AI literacy vary by sector and context. Chee et al.~\cite{chee2024competency} present a framework that varies by learner groups, emphasizing basic AI knowledge and ethics for K-12, data understanding and problem-solving for higher education, and data interpretation and AI-based decision-making for the workforce. 


These frameworks have offered actionable design considerations for integrating AI literacy into school curriculum and assessments \cite{yim2025artificial,casal2023ai}. Empirical studies also demonstrate that age-appropriate, hands-on curricula can foster conceptual understanding and with measurable improvements \cite{su2022meta}. 
Beyond the classroom, families and especially parents, can play a significant role in supporting children and adolescents' AI literacy development. Many children's early interactions with AI occur at home (e.g., smartphones and voice assistants), where parents mediate these experiences \cite{andries2023alexa}. Empirical work by Druga et al.~\cite{druga2022family} highlight diverse ways parents support children's understanding of AI and that families act as a ``third space'' for active AI learning. Nonetheless, parents can encounter several challenges during the process. Many parents report limited understanding of how AI systems function, which undermines their confidence in scaffolding children’s learning \cite{druga2022family, sun2024exploring}. The rapid rise of generative AI undoubtedly expanded children's everyday access to AI-empowered tools, with a quarter of teens reporting using ChatGPT for schoolwork in 2024 and awareness reaching 79\% by January 2025 in a U.S. survey \cite{SidotiEtAl2025_ChatGPTTeens}. These shifts create new challenges for parents to provide age-appropriate guidance and supervision. Critical gaps still remain in our understanding of how to support families in fostering AI literacy. Exploring how parents conceptualize AI literacy across developmental stages and how they manage AI use at home accordingly will provide valuable insights for designing family-child partnerships in AI learning.

\subsection{AI for Self-Directed Learning}

Self-directed learning (SDL)~\cite{knowles1975self} is a process in which individuals take the initiative and responsibility for their own learning endeavors, from diagnosing their learning needs to identifying resources, implementing learning strategies, and evaluating outcomes. Unlike traditional, teacher-led instruction in formal school settings, which is often highly structured and uniform, SDL is characterized by learner autonomy, intrinsic motivation, and goal ownership~\cite{tough1979adult}. It is closely related to, and often encompasses, self-regulated learning (SRL), which refers to the cognitive, metacognitive, and motivational processes that learners use to control their own learning~\cite{zimmerman2002becoming,panadero2017review}. While SRL highlights internal regulation mechanisms, SDL focuses more broadly on learner agency and environmental navigation, making it highly relevant for home learning with AI.

Recent advancements in artificial intelligence, particularly generative AI, have introduced powerful tools to scaffold and enhance the SDL process~\cite{younas2025comprehensive}. Research indicates that AI systems are increasingly being used to support learners across the entire learning cycle. AI tools like ChatGPT can assist learners in setting goals, locating resources, and creating personalized study plans~\cite{lin2024exploring,chun2025planglow,liu2025enhancing}. They serve as interactive partners for brainstorming ideas~\cite{wang2024understanding}, practicing conversations for language learning~\cite{dizon2024chatgpt,li2024reconceptualizing}, and acquiring technical skills like programming~\cite{ouaazki2024generative,puttajanyawong2025utilizing}. Furthermore, AI-powered e-tutors and chatbots can provide personalized, adaptive feedback and real-time support, which studies have linked to increased learner motivation, engagement, and overall performance~\cite{maphalala2025exploring,li2025impact}. 
Despite these promising applications, current research highlights persistent challenges and reveals a critical gap. A primary concern is the potential for learners to become over-reliant on AI, alongside issues of information accuracy and reliability, which necessitates that learners develop critical evaluation skills~\cite{lin2024exploring,ouaazki2024generative,zhang2025research,dizon2024chatgpt}. The effectiveness of these tools often depends on the user's ability to craft effective prompts and the presence of teacher support, indicating their successful use is not automatic~\cite{lin2024exploring,garg2024analyzing,wu2024unlocking}. Critically, existing research on AI for SDL focuses almost exclusively on the individual learner, predominantly in higher education or adult learning settings~\cite{wang2024understanding,esiyok2025acceptance,biyiri2024chatgpt}. While some work anticipates applications for K-12 students, it does so from the perspective of a tool for individual use~\cite{ali2023supporting}. This leaves a significant gap in understanding how AI-powered learning tools are adopted and used within the family context. The role of caregivers in mediating children's use of AI, guiding their learning, and mitigating risks like dependency remains largely unexplored. Our study addresses this gap by investigating family perspectives to inform the design of responsible AI that can effectively and safely support self-directed learning for K-12 children at home.

\subsection{Family-Centered Design for Emerging Technology}

Safeguarding children’s interactions with technology has long been a significant concern among parents, educators, and researchers. The theoretical foundation for many family-centered design approaches lies in parental mediation theory, originally developed in the pre-Internet era to address parental strategies around children’s television consumption. Studies by Nathanson and Valkenburg formalized three core mediation strategies: restrictive mediation (setting rules), co-viewing (watching together), and active mediation (discussing content) \cite{nathanson1999identifying,valkenburg1999developing}. With the advent of the Internet, parental mediation theory evolved to balance minimizing online risks with maximizing digital opportunities for children \cite{livingstone2008parental,mesch2009parental}. Notably, restrictive mediation came to include technical controls, such as filtering and screen-time limits, while active mediation remained grounded in dialogue, and co-viewing shifted to more surveillance-like online monitoring \cite{clark2011parental,nikken2014developing}. Recent empirical work has sought to situate parental mediation strategies within broader family dynamics. For example, Ren and Zhu~\cite{ren2022parental} examined how parenting styles influence the effectiveness of mediation strategies, finding that active mediation often aligns with supportive parenting approaches and is positively associated with learning-related internet use.

Within the HCI and design literature, scholars have increasingly called for a family-centered perspective in the development of emerging technologies. Cagiltay et al.~\cite{cagiltay2023family} argue for embedding family systems theory into child-robot interaction design, highlighting that technology adoption and long-term engagement depend on aligning designs with family routines, relationships, and values. Similarly, studies of parental control apps emphasize how design features such as transparency, granularity of control, and mechanisms for communication between parents and children shape both parental experiences and children’s sense of autonomy \cite{wang2021protection}. Moreover, there is growing recognition that emerging technologies should move beyond risk mitigation toward fostering children’s digital autonomy and emotional well-being. Wang et al.~\cite{wang2023empower} propose design mechanisms, ranging from scaffolding and peer support to digital playgrounds, that cultivate autonomy rather than merely enforcing restrictions. 
In the era of AI, family-centered design faces new challenges such as mismatched perceptions between parents and children regarding AI risks--parents often focus on privacy and misinformation \cite{zhang2025exploring,yu2025exploring}. Reports of unsafe behaviors in therapeutic chatbots and growing parental concerns about AI in children’s apps further reveal gaps in safeguards and oversight \cite{Chow2025}. Yet existing theories rarely account for AI’s relational and persuasive roles in family life, emphasizing the need for updated frameworks and design principles that balance child autonomy, family communication, and ethical protections in AI-mediated environments \cite{szondy2025artificial}.

\section{Formative Study}

We first conducted a formative study to explore parents’ perspectives on children’s use of AI, focusing on their current practices, concerns, and expectations, which helped us formulate questions and procedures for the formal study.

\subsection{Method} 

We recruited 10 parents from 7 Chinese families (7 mothers, 3 fathers) through prior contacts and snowball sampling. Participants were eligible if they (1) had a child enrolled in elementary or junior high school and (2) were interested in using AI tools with their children. Participation was voluntary and uncompensated. 

We conducted one-on-one semi-structured interviews covering: 
(1) family background (parent demographics and children’s grade levels),
(2) parents' interest and perspectives on using AI with their children (scenarios, benefits, challenges), 
and (3) attitudes toward children’s AI use.
Interviews lasted 27-50 minutes and were audio-recorded after consent. 
Two authors conducted an inductive thematic analysis~\cite{braun2006using}, iteratively coding the transcripts and grouping codes into themes. All authors reviewed and refined themes in weekly meetings.

From this process, we derived three key insights into parents' concerns about children's use of AI, their strategies for addressing these concerns, and their expectations for children’s gradual engagement with AI. These insights informed the design of our focus group materials.

\subsection{Key Insights}
\label{formative_findings}

All participants have used AI to support children's learning, and their concerns regarding children's use of AI were closely related to the learning context.

\paragraph{\textbf{Parents' Concerns about Children's Over-reliance on AI}}

All participants worried that children might become overly reliant on AI, using it as a shortcut that replaces their own thinking. 
One participant explained, \textit{``my child gets an answer from AI easily and instantly, then just copies AI’s answer without thinking it through.''}

Participants noted that while AI helps children finish tasks (e.g., homework), it does not foster deeper understanding. This led to a perceived gap between polished homework completed with AI and children’s actual understanding, particularly in exams: \textit{``even those who don't do well in exams end up turning in homework that's completely correct.''} 
Some parents observed children's grade improvements with AI usage, but at the cost of increased dependency, as children often \textit{``check if AI’s solution matches mine.''}

\paragraph{\textbf{Parents' Involvement in Children's AI Use: Monitoring, Reinterpreting, and Guiding}} 
To mitigate these concerns, all participants were actively involved in children's use of AI by monitoring child-AI interactions, reinterpreting AI's answers, and guiding children on how to use AI. 

Participants rarely allowed children to use AI for learning independently, noting, \textit{``The whole process [of using AI] takes place under my supervision.''} They explained that without monitoring, children's reliance on AI could deepen, and they might use electronic devices for entertainment. 
%
Alternatively, some parents use AI themselves by first interpreting its outputs and then re-explaining it to their children.

Participants also showed children what they considered appropriate ways of using AI for learning. They viewed AI as a source of new ideas and perspectives, instead of simply an answer machine for specific homework questions. For example, some asked AI to provide multiple solutions to math problem, encouraging children to think from different angles. Others generated several example when tutoring composition, then reviewed them with their child, using questions to guide the child in developing their own ideas.

\paragraph{\textbf{Age-related Progressive Use of AI}}
Parents expected their involvement and restrictions on children's use of AI to decrease as children grow older, with children's access to AI progressing gradually by age. 
One parent explained, \textit{``Primary school children are still developing cognitively. They cannot yet evaluate AI's outputs due to their limited knowledge base and cognitive abilities. In contrast, middle and high school students are at different developmental stages, and their capacities evolve accordingly [so they can use AI more effectively].''}

This progression often begins with no direct access (parents use AI and then teach children), and gradually moves toward independent use once children showed greater self-discipline. Participants agreed that children will eventually use AI on their own, noting, \textit{``since we live in a digital age, it's unrealistic to expect children not to use AI.''}

\section{Method}


\subsection{Participants} 
We recruited 13 parent-child pairs from 13 Chinese families through prior contacts and snowball sampling. Each pair consisted of (1) a child enrolled in elementary or junior high school, and (2) at least one parent with prior experience using AI tools. We encouraged both parents to participate when they each had such prior experience. 
%
In total, 28 individuals participated: 15 adults and 13 children, including 8 mother-child pairs, 3 father-child pairs, and 2 groups with both parents and one child. Participant demographics are summarized in Table~\ref{table_demo}. Each household received 100 RMB ($\approx$15 USD) if one parent participated, or 200 RMB ($\approx$30 USD) if two parents participated.

\begin{table}[htbp]
\centering
\caption{Family Demographics and Children’s Electronic Device Usage}
\label{tab:family-columns}
\resizebox{\linewidth}{!}{%
\begin{tabular}{llllllllllll}
\toprule
& \textbf{Parent} & \textbf{Parent} & \textbf{Parent} & \textbf{Parent} & \textbf{Parent} & \textbf{Child} & \textbf{Child} & \textbf{Child} & \textbf{Child} & \\
\textbf{Family} & \textbf{ID} & \textbf{Role} & \textbf{Age} & \textbf{Education} & \textbf{Occupation} & \textbf{ID} & \textbf{Age} & \textbf{Grade} & \textbf{Gender} & \textbf{Personal Devices} \\
\toprule
H1  & H1M & Mother & 36--40 & Bachelor  & Teacher      & H1K & 9  & G4 & M & Tablet \\
\hline
H2  & H2M & Mother & 36--40 & Associate & Operator     & H2K & 13 & G8 & M & Tablet \\
\hline
H3  & H3M & Mother & 36--40 & Bachelor  & Technical Staff   & H3K & 9  & G4 & M & Tablet, Laptop \\
    & H3D & Father & 36--40 & Bachelor  & Telecom Engineer  &     &    &    &   &                 \\
\hline
H4  & H4D & Father & 41--45 & PhD & Scientist & H4K & 11 & G6 & M & Laptop \\
\hline
H5  & H5M & Mother & 36--40 & Master & Technician & H5K & 7 & G2 & F & No Personal Devices (use parents') \\
\hline
H6  & H6D & Father & 36--40 & PhD & Professor & H6K & 12 & G7 & M & Smart Phone, Laptop, Tablet \\
\hline
H7  & H7M & Mother & 31--35 & Bachelor & Teacher & H7K & 9 & G4 & M & Tablet \\
\hline
H8  & H8M & Mother & 36--40 & High School & At-home Mom & H8K & 7 & G2 & M & Smart Phone, Tablet \\
\hline
H9  & H9M & Mother & 41--45 & Bachelor & Teacher & H9K & 10 & G5 & F & Tablet, Smart Watch \\
\hline
H10 & H10M & Mother & 36--40 & Bachelor & Human Resources & H10K & 11 & G6 & M & Tablet \\
\hline
H11 & H11M & Mother & 36--40 & Bachelor & Office Staff & H11K & 12 & G6 & F & Tablet \\
\hline
H12 & H12D & Father & 36--40 & PhD & Professor & H12K & 11 & G6 & F & Tablet, Laptop \\ \hline
H13  & H13M & Mother & 36--40 & Associate & Sales        & H13K & 12 & G7 & F & Learning Tablet* \\
    & H13D & Father & 36--40 & Associate & Electrician  &     &    &    &   &                                   \\
\bottomrule
\end{tabular}%
}
{\raggedright \small* Learning tablet is an AI-powered learning device for students that combines tablet hardware with educational software. It offers personalized learning support (e.g., problem-solving guidance, error analysis, and personalized study plans) and allows parents to track learning progress~\cite{iFlytek2025}.
\par}
\label{table_demo}
\end{table}

\subsection{Study Materials}

Based on key insights drawn from the formative study (Section~\ref{formative_findings}), we outlined four possible phases of how children use AI: (1) \textit{Asymmetric Familiarity}, parent is familiar with AI, child is not (or vice versa); 
(2) \textit{Joint Engagement}, parent and child co-use AI; 
(3) \textit{Guided Use}, child uses AI under parental monitoring; 
(4) \textit{Independent Use}, child uses AI independently without parental involvement.
Here, we use child to denote the parent-child relationship, regardless of the child's age.

We illustrated these phases on a Miro board~\cite{miroboard}. 
The Miro board shows the names of the phases and a timeline connecting them. Within each phase, we provided (1) a description of the phase and its objectives, 
(2) examples of AI tools or functions that might be relevant, 
(3) examples of situations in which children’s use of AI could be seen as undesirable by parents, 
(4) examples of ways parents and children might negotiate or resolve such situations,
and (5) examples of potential design features or expectations for AI tools to help address these challenges.  
Note that these examples were not intended to represent the ``correct'' answers or categories. Instead, they were prompts designed to support participants in recalling their own experiences and to facilitate open-ended discussion.

We also developed focus group protocols for warm-up activities, to help participants understand the illustration on Miro board, and to evaluate children's current AI competence. 
To further probe family conversations around AI and to explore both parents’ and children’s understandings of AI literacy, we designed a ten-question instrument (See Appendix~\ref{tenquestion}) adapted and shortened from existing survey scales \cite{long2020what,touretzky2019ai4k12,unesco2021aicompetency}, for the length of our study. The AI literacy items addressed conceptual knowledge, everyday awareness, capabilities and limitations, simplified mechanisms, and ethics/safety, while the SDL items were reframed to capture how children envision using AI across phases of self-directed learning, such as readiness and goal setting \cite{knowles1975self,garrison1997self}.

\subsection{Study Procedure} 
We conducted focus groups with parent-child pairs from 13 households via online video conferencing platforms in the summer of 2025. Each session involved one household at a time, with one or two parents and their child, and lasted 45-90 minutes.
At least two researchers attended each session, where one facilitated the discussion by asking questions, while the other shared the screen to display the Miro board and took notes based on participants' insights.
With participants’ consent, sessions were audio recorded, and the Miro board screen was video recorded.

During the focus groups, all questions were directed to all the participants, including both parent(s) and their child. Researchers encouraged and reminded parent(s) to engage their child in the discussion and invited children to share their perspectives. The focus groups followed the steps below.

First, we introduced the study objectives and collected demographic information, including children's age and grade level, access to personal electronic devices, parents' and children's experiences using AI, and any school or family regulations regarding how children use AI.

Second, we presented the content on the Miro board. We described the predefined phases of children's AI use and asked participants for their feedback. 
They were encouraged to select, modify, remove, or reorder the phases. We also asked them to indicate the expected child age or grade associated with each phase, as well as conditions under which children might transition to the next phase.

Third, we explored each phase determined by participants in greater detail. Participants shared AI tools or functions that children have used, wish to have, or prefer to block. They also discussed the benefits and limitations of these tools or functions.

Next, we asked children the list of ten questions (See Appendix~\ref{tenquestion}) to understand their AI literacy. 
Children first responded individually while parents remained silent to avoid influencing their answers. After completing the full set, parents and children engaged in joint discussion, during which parents could review and elaborate on the child’s responses. Researchers did not intervene except when explicitly invited to clarify. Importantly, neither scoring criteria nor results were shared with families, as the questions were not used as an assessment but rather as prompts to reduce performance pressure and encourage open reflection and dialogue.

Finally, participants discussed situations in which children's use of AI tools was considered undesirable by parents.
Participants reflected on past situations and described how parents and children had reached consensus or which strategies parents had adopted to mitigate conflict. They were also encouraged to envision potential future situations, along with possible solutions and AI tools or functions that might alleviate such conflicts.

\subsection{Data Analysis}
We conducted a fully inductive thematic analysis of focus group discussions with 13 families \cite{braun2006using,braun2019reflecting}. Three researchers independently coded transcripts from the first four families, each developing preliminary themes. These were then compared and discussed to build an initial shared foundation. As additional families participated in the study, we iteratively added, merged, removed, and refined codes and themes through cycles of independent viewing followed by weekly collaborative discussion, including three additional families in each round. This iterative process of divergence and convergence ensured that the final themes reflected both individual insights and collective consensus, resulting in a set of themes grounded in the data and agreed upon by all three researchers.

\section{Findings}

This section presents findings on parental perspectives of children’s AI use. Parents considered children’s progression with AI in relation to screen time, self-directness, and knowledge growth. They also described monitoring practices shaped by AI’s pervasiveness in daily life, as well as their limited understandings of AI literacy and opportunities to develop it with their children.

\subsection{Factors Shaping the Progression of Phased AI Levels}
\label{finding_factors}
Despite variations in parents' approaches to managing AI, three factors are commonly considered by parents when they evaluate children's progress along the phases of AI usage, as detailed below. 
\vspace{-0.2cm}
\begin{figure}[htb]
    \centering
    \includegraphics[width=\linewidth]{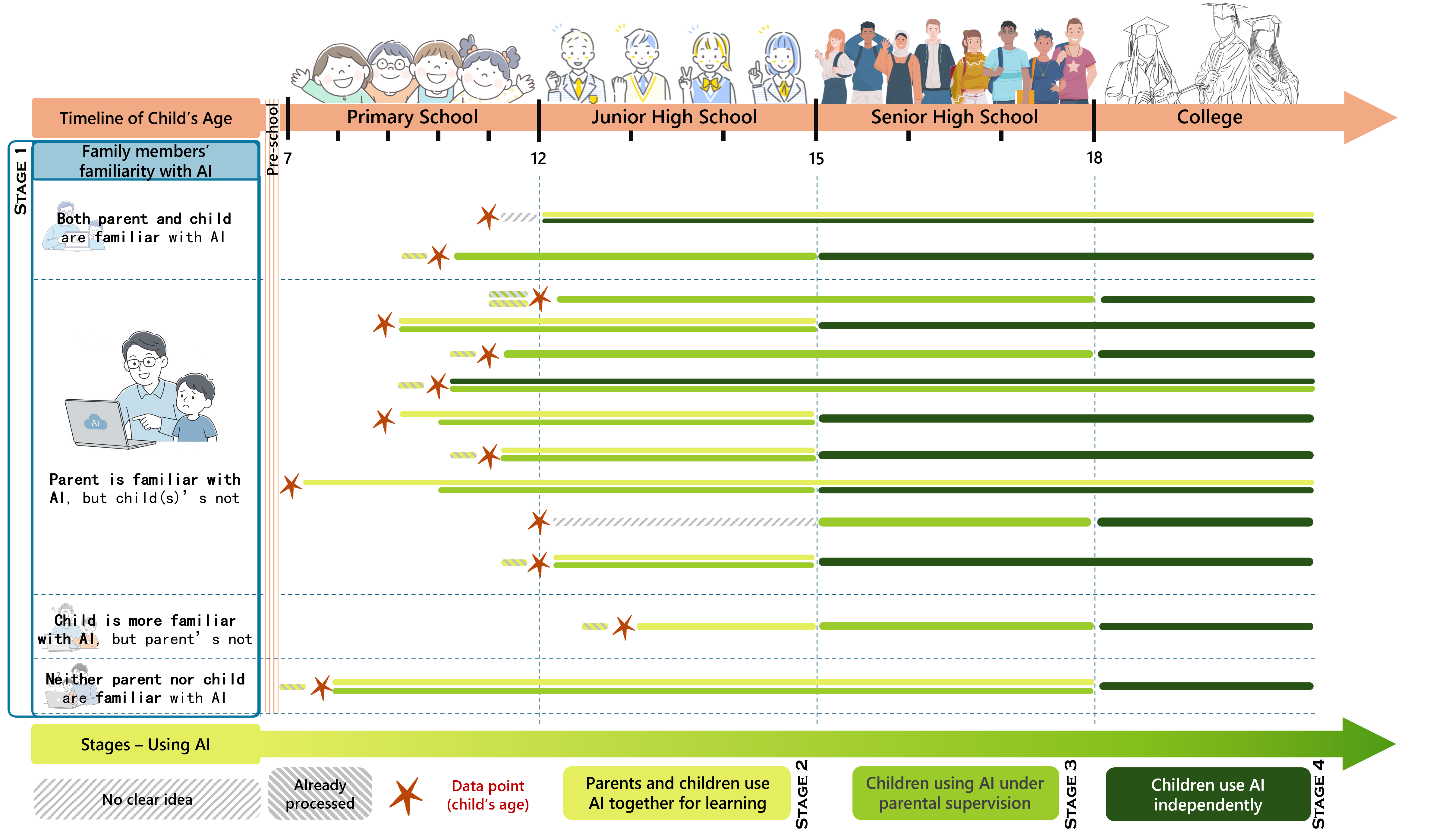}
    \vspace{-0.2cm}
    \caption{\textbf{Staged family plans for children’s AI use}. Each horizontal bar represents one family’s planning pathway; each five-pointed star locates the child’s current age. The left side shows Stage 1: family (parent and child) familiarity with AI in four groups. The right side maps subsequent stages along the age timeline: Stage 2 (light yellow-green) = parent and child use AI together for learning, Stage 3 (medium green) = child uses AI under parental supervision, and Stage 4 (dark green) = child uses AI independently. Dashed grids and school-level markers (Primary/Junior High/Senior High/College) align age anchors (e.g., 12, 15, 18). Diagonal hatching denotes periods with no clear plan, and light-gray shading indicates stages already completed. The visualization highlights both convergence, such as delaying independence until after primary school and clustering transitions around junior high, and divergence, as families differ widely in when and how they envision autonomy, with “independent” use sometimes still bounded by parental oversight. (The lines are not ordered by Household ID)}
    \label{fig:data-points}
\end{figure}

\subsubsection{Overview on Parental Pathways and Expectations for Children’s AI Use}
Parents' perceptions of children's phased stages of AI usage, as reported in the study, are visualized in Figure~\ref{fig:data-points}. In general, parents find the proposed four phases of children's AI usage reasonable and applicable. There are some minor adjustments based on different perceptions of AI and family situations. For example, regarding \textit{Asymmetric Familiarity}, H9M commented that the condition that the parent is familiar with AI and the child is not should not exist, since parents would always seek to learn more about a technology before feeling comfortable and providing its access to children. Concerning \textit{Joint Engagement} and \textit{Guided Use}, H10M believed the two phases would be simultaneous instead of sequential, as parents who use AI in collaboration with children are often monitoring them at the same time. 

The staged timelines reveal both points of convergence and divergence in how families envision children’s AI use across ages. Across households, there was broad agreement that independent AI use should be delayed until at least the end of primary school, with most families expecting joint engagement or strict parental oversight during ages 7--9. Around age 12, coinciding with the transition to junior high, many families marked a natural turning point, shifting from co-use toward supervised use, suggesting consensus that early adolescence is an appropriate stage for increased access but not yet full autonomy. Beyond this point, however, trajectories diverged. Some families anticipated independence as early as mid-junior high, while others extended close supervision through senior high, delaying autonomy until college. 

Several families also left transitional years undefined, reflecting either uncertainty about when children would be ready or a lack of clarity in parents’ own expectations. A few pathways further showed overlapping phases (e.g., joint use and supervision occurring simultaneously), suggesting that families do not always see these stages as strictly sequential. Notably, even when parents described ``independent'' use, this rarely meant full autonomy; rather, it implied a loosening of direct supervision within boundaries parents still considered appropriate. 

Together, these similarities and variations underscore differing parental interpretations of children’s readiness and the staging of AI use, which we elaborate on in the sections below.

\subsubsection{Compliance with Screen-Time Rules as a Benchmark for AI Use}
\label{screentime}

Parents in our study used screen time as a measure for children’s exposure to AI and as a benchmark for evaluating their readiness for further engagement. They believed that if children could comply with screen-time restrictions, then they were trusted with greater autonomy in using AI tools and would be allowed for more exposure to AI. As a result, screen-time management became a criterion for evaluating children’s progress toward more independent AI use. H10M, for instance, described a shared consensus among parents. 

\begin{quote}
    \textit{``Some friends and I, all parents of children with different academic performances... share the same view that children should spend less time on electronic devices. This past summer vacation made it clear to us that once you let them play, it’s hard for them to stick to agreed-upon time limits. Most of the time, they use the devices for games, and games are hard to stop abruptly.''} (H10M)
\end{quote}

Her example illustrates a core challenge: the issue is not only about limiting device time but also about ensuring children follow agreed-upon rules. Therefore, even parents who hold positive views about AI often regard its most common mediums (e.g., smartphones, tablets, or laptops) as having a negative impact, especially for younger children. This perception consequently leads to children's limited interactions with AI, which we elaborate in Section~\ref{finding_monitor}.

H9M reinforced this point by observing that children can quickly redirect device use toward unintended activities. She indicated that \textit{``It's impossible to avoid [undesired usage]... even if you're sitting right beside them, the moment you look away, they’re already using the tablet for something else.''}

Overall, the management of children's AI use was inseparable from parents' regulation of screen time, with compliance considered as a visible indicator of children’s discipline. However, this approach remained an external form of control, in contrast to children's development of internal regulation, which we examine in Section~\ref{finding_sd}.

\subsubsection{Self-Directness as the Key to Autonomy}
\label{finding_sd}

Children's preparedness for SDL has been highlighted by parents throughout the study as a core criterion for determining the level of autonomy children should have in AI usage. Here, self-directness refers to children's ability to set personal goals, make decisions, take initiative, and guide their own behavior and learning without over-reliance on external control. Parents described this capacity through concrete examples in AI use, specifically regulating time, making sound judgments, and preserving independent thinking. These three dimensions serve as benchmarks for evaluating children’s readiness for greater exposure to AI.

The first dimension is time management. It echoes parents' concerns about screen time (Section~\ref{screentime}), reflecting a need for children to develop skills in regulating their use of technology. Children are aware of their parents' expectations to self-manage time, yet they also recognized that they often struggle with this capacity, as mentioned by H1K. 

\begin{quote}
    \textit{``Every day after lunch and dinner, I watch for half an hour, then I have to turn it off. I always keep an eye on the clock, but sometimes, when I’m watching an exciting part of a cartoon and the time’s up, I don’t want to stop. My dad always tells me that self-discipline means turning it off when time's up.''} (H1K)
\end{quote}

The second dimension is sound judgment. Parents emphasized that autonomous AI usage requires children to distinguish between beneficial and harmful content. Younger children, in particular, were seen as more vulnerable and would be, as H3D explained, \textit{``easily influenced by harmful content.''} 
In such cases, parents were \textit{``not recommended for child to use AI independently''} and instead, anticipated the child to only access material that \textit{``benefits his personal growth and doesn’t distort his values.''} (H3D) H10M echoed this view, adding that without judgment, AI could become another form of entertainment (e.g., mobile games) rather than a useful aid for children (e.g., a learning tool).

The third dimension of self-directness is independent thinking. Parents expected children to preserve their independent thinking ability when using AI. Specifically, children should not use AI as a shortcut to complete their learning tasks or bypass their thinking process:
\begin{quote}
    \textit{``I think children sometimes struggle with self-control. I hope he will use pen and paper to work through problems seriously on his own. For example, when teachers assign homework, I want him to think it through and summarize the answers, rather than relying on AI to do it for him.''} (H4D)
\end{quote}
H4D articulated that childhood is an important stage for developing summarizing, analyzing, and problem-solving skills. Over-reliance on AI, however, was seen as undermining this development when children delegate such tasks to AI instead of practicing them.  
Parents therefore believe that children with stronger self-directness would be less likely to rely on AI and would \textit{``think on their own before resorting to AI for a solution''} (H5M). 

Despite agreement on the necessity of self-directness, there are variations in when and how it could be cultivated. Some parents believed children would gain self-directness as they advance to the next level of education. For example, H1M expected her child to search for study-related materials on a tablet \textit{``by the time he reaches middle school, or maybe fifth or sixth grade of elementary school.''}
However, H4D found it difficult to judge based on age, and emphasized guiding children to improve their abilities. H9M also noted that self-directness is a complex characteristic shaped continually by parents, family, and the overall environment:
\begin{quote}
    \textit{``For children whose parents have never been exposed to modern technology, they often become very curious when they first encounter it. But if parents set limits early -- allowing kids to explore, play in moderation, and gain a basic understanding of these things -- then children are less likely to become so addicted.''} (H9M)
\end{quote}

In addition to variations among families, H9M highlighted that controlled exposure at an early stage may prevent overuse and addiction later on. Similarly, H7M argued that strict parental regulation may provoke rebellion as children gradually develop a sense of autonomy. Instead, she described self-directness as a process of building mutual trust, in which children follow instructions and parents respond by granting them greater autonomy. Overall, parents regard self-directness as a prerequisite for safe and effective AI use and a skill developed through guidance and trust.

\subsubsection{Knowledge Base Contributing to Constructive AI Interactions}
Conceptual knowledge constitutes an important aspect of AI literacy. Parents in the study agree that a certain level of understanding in AI is necessary for children to have constructive interactions with AI: 
\begin{quote}
    \textit{``I think she needs to learn more about what’s behind it (AI) like its principles, how it works, and its reasoning mechanisms. That would help her build understanding and develop her own judgment. I think she should know what AI is good at and what it isn’t, because to use it effectively, she needs to understand both its strengths and limitations.''} (H5M)
\end{quote}
As H5M stated, the effective use of AI is dependent on the strengths and weaknesses of AI, and children need prior knowledge to decide the appropriate usage scenarios. H4D supplemented that AI does not always give the correct answer, which requires children to use their knowledge, in collaboration with parents, to validate some of the answers. This may seem difficult for young children, but H3D believes AI is unlikely to make errors on elementary school-level questions, and it is in middle school or beyond that children need to be mindful and not to take AI's output for granted.

Parents also considered safety knowledge as part of the broader foundation needed for AI use. For example, H7M described teaching children about online scams and ensuring that only adults manage payment-related accounts. Although these efforts were framed as general internet safety, parents considered such caution transferable to AI, since interacting with AI systems also requires awareness of risks and responsible use. 

Beyond safety considerations, parents emphasized that the effectiveness of AI use depends on a child’s broader intellectual readiness. For example, H3D assumed that \textit{Independent Use} should not begin until high school, where the child \textit{``knows what he wants to learn, has developed his own values, his own ideas, and knows what questions he wants to ask.''} H13M further noted that AI interactions should be intellectually stimulating, as intelligent AI systems can adjust to a learner’s ability by presenting more complex challenges. This opinion connects a child’s knowledge base with the extent of AI exposure, since the system’s benefits depend on the child’s readiness to engage at higher levels.

\vspace{2mm}

In summary, from parents' perspectives, children's progression in phased AI usage is not simply about learning what AI is and how to use AI. Instead, it is often considered and evaluated within screen time management, self-directness development, and knowledge growth, topics that are indispensable in parenting. Nevertheless, the rapid advancement in AI and pervasive AI integrations in daily life present new challenges and motivate nuanced monitoring practices. 

\subsection{Parental Monitoring Practices in the Context of Pervasive AI} 
\label{finding_monitor}

Similar to our findings in the formative study (Section~\ref{formative_findings}), parents from eight household (H1-3, H7-9, H11, and H13) expressed concerns that children's excessive reliance on AI may impede the development of their learning skills. 
In response to these concerns, parents attempt to monitor and regulate children's AI usage by focusing on scenarios (non-learning contexts) and platforms (app-based restrictions). However, their strategies reveal a tension between intended restrictions and the seamless integration of AI into digital life. In this section, we uncover the limitations and opportunities of these parental monitoring practices.

\subsubsection{Constraining AI to Learning-Only Functions While Rejecting Broader Uses} 
\label{finding_nonlearning}

Building on the general concerns over screen time, parents from nine households (H1M, H3D, H4D, H5M, H7M, H9M, H10M, H11M, H13M) highlighted a related but distinct issue: the need to restrict AI when its use is disconnected from learning purposes. While some parents tolerate or even encourage AI for educational games or video-based learning, they consistently expressed reservations about entertainment-focused activities such as casual gaming, social media browsing, or video streaming. H13M explained, \textit{``Learning English vocabulary while playing a game is great, but if it’s just pure gaming without any learning, I think most parents would probably reject it.''} She further described and promoted a gamification-based learning system favored by her child (H13K), where children earn points by answering questions correctly and redeem those points for mailed rewards or by caring for a virtual pet.

However, parents repeatedly noted the difficulty of separating learning from play. As H10M put it, \textit{``The tablet is already a rare entertainment tool. Adults might think of AI as versatile, but children just treat it as a way to play games.''} Even when entertainment apps are disabled, children often repurpose AI for game-related uses. For instance, H10M shared that her child frequently asked AI chatbots gaming questions: \textit{``When he talks with DeepSeek, all his questions were things like `How do you level up in PUBG Mobile?' or `How do you beat the boss?' those very specific game-related questions.''} 
These concerns led some parents, such as H4D and H7M, to suggest AI companies design clearer boundaries for educational tools. As H7M proposed, \textit{``simply block entertainment-related features such as casual chatting.''}

Parents also described two main oversight strategies. The \textit{first} is limiting entertainment-oriented AI apps while allowing greater flexibility for learning activities. For instance, H3D restricts leisure-oriented AI tools (e.g., RedNote) to 15 minutes per day, while learning apps are either unrestricted or permitted for much longer periods. The \textit{second} strategy is co-using AI, where parents accompany children during AI interactions.

\begin{quote}
    \textit{``If I notice that he often asks off-topic questions when using AI [alone] to look up homework problems, then I use AI together with him. I'm the one doing the searching, and I use the AI content to guide him.''} (H10M)
\end{quote}

Even when children reach a stage where they can use AI independently, some parents remain cautious. As H13M noted, \textit{``it's really hard to believe children would only use AI for studying and avoid everything else. Even adults can’t always be self-disciplined.''}

\subsubsection{App-Based Oversight Amid AI's Infrastructural Embedding} 
\label{appbased_oversight}

App-based control is currently adopted by most parents to track and restrict children's exposure to AI. Children may not be allowed to download standalone AI apps (e.g., banning ChatGPT), or they may be permitted to use these apps under parental supervision or with screen-time limits. 

However, as AI becomes increasingly embedded into daily life, it has shifted from being just an app-based tool to operating as the underlying infrastructure supporting a wide range of applications. Monitoring app usage, though tangible, becomes inherently limited and fails to address AI's infrastructural invisibility. Consequently, children can access AI even without parental awareness.

Parents themselves sometimes recognize this limitation. H10M noted that although they banned the app, the child could still use the browser to search for related content. H9M gave an example of her child who \textit{``is interested in interacting with the voice-based navigation bot in the car, almost like arguing with it, and she can keep at it for quite a while.''}
These examples demonstrated that app-based restrictions cannot contain what is already infrastructurally integrated.

Participants reported that even with some app bans, children still learned about or accessed AI through: 
(1) electronic devices like smartphones, laptops, and tablets (H2M, H3D, H7M, H9M, H13M);
(2) online video platforms like TikTok and YouTube (H2M, H6K, H7M);
(3) AI chatbots like DeepSeek, ChatGPT, and Cici (H3D, H7M);
(4) productivity software like Google Docs (H6K); 
or (5) in-car voice assistants (H9M). 
Among these families, H2M and H3D had no clear idea how and when their children were exposed to AI. It was during the study that H13M first realized her child had bypassed parental monitoring and used AI on smartphone to generate outlines for school compositions.
Children also described these workarounds directly. One child (H6K), despite bans from both school and parents, explained how AI appeared in platforms not explicitly labeled as ``AI apps'':
\begin{quote}
   \textit{``Google Gemini like summing up stuff when I do Google search and then probably like the YouTube algorithm or something. One time I was on YouTube and then I saw a video, it was about why AI can't draw hands, right? And then I learned a lot from that.''} (H6K)
\end{quote} 

To mitigate this challenge, H4D emphasized that he did not want his child to use nowadays all-in-one AI, such as those \textit{``offering both learning and gaming functions.''}
Instead, he preferred earlier generations of AI apps that \textit{``focus on one specific function, so that in many cases there’s no need for `monitoring' at all.''}

\subsubsection{Content Awareness of Child-AI Interaction as Parenting Practice}

Beyond restricting access, some parents have been monitoring the content of children's interactions with AI. They consider AI logs as a risk-detection mechanism to identify unexpected use and an opportunity to learn children's interests and developmental needs.

A major reason for content monitoring is concern about children's exposure to non-age-appropriate material, reported by six parents (H1M, H3D, H7M, H8M, H9M, H11M). 
They mentioned that the platforms children use to access AI (Section~\ref{appbased_oversight}) often lack robust age-rating systems.
As a result, children may encounter negative influences such as foul language or inappropriate behaviors.
For younger children, parents like H1M are worried that \textit{``they just imitate it even without understanding what it means.''}  
H9M extended this concern to the broader design of AI systems, pointing out that these tools are developed by humans whose values and intentions remain opaque to users: 
\begin{quote}
    \textit{``I don't know who designed these AI tools, what values they hold, or what ideology they follow. Could there be hidden modes, like secret levels in video games, that parents don’t know about? What if the AI exposes my child to violence or extreme, biased ideas without me realizing it? How can I tell whether this AI app is truly suitable for my child’s growth and mental health? That's what makes me anxious.''} (H9M)
\end{quote}

For these reasons, parents monitored content not only to detect risks but also to counsel their children when needed. Importantly, they emphasized that such monitoring could not be delegated entirely to AI systems. Logs and time records may provide useful signals, but parents' judgment and involvement are essential. As H3D explained, while AI monitoring may ease parental effort, it ultimately \textit{``feels like parents are just handing the kid over to a robot, or to some `thing'.''} Instead, through reviewing content of child-AI interaction, parents sought to interpret children’s emerging interests and identify challenges they were facing. H3M elaborated: 
\begin{quote} 
    \textit{``Parents still need to exercise their own judgment. For example, if the child searches for something, you have to evaluate what it is used for. And then you need to guide them in response. If they’re searching sensitive terms like dating, then as a parent you definitely need to pay attention. That's not something you can just leave entirely to AI.''} (H3M)
\end{quote}

Participants further revealed two approaches to content monitoring: \textit{building trust with children} and \textit{avoiding over-monitoring}. First, parents positioned themselves as partners rather than overseers. They built rapport by acknowledging children’s interests and engaging in open conversations, such as reminding them that \textit{``what they see online isn’t always the full truth''} (H5M). These efforts were framed as long-term trust-building, particularly important in the early years, as H5M put it, \textit{``monitoring combined with reasoning makes sense when children are young, but strict control isn’t very effective once they get older.''}
Second, parents articulated the importance of avoiding over-monitoring, which they viewed as potentially harmful to the parent-child relationship (H3D, H2M, H3M, H4D, H7M). For example, rather than reading entire logs, some parents preferred keyword checks. 
H3M also described a balance of awareness without constant interference. She believed that parents do need to supervise, but \textit{``can’t interfere with everything you see or monitor every detail, otherwise the child’ll push back even more. You need to be aware, but that doesn’t mean you always step in.''}

\subsubsection{Parental Authority over Children’s Voices} Although the focus groups included parent-child pairs, the conversations were dominated by parents, with children’s perspectives rarely acknowledged. Children did not comment on the staged “phases” of AI use; parents presented these plans as settled decisions, and children did not challenge or reinterpret them. When prompted to engage their children, most parents directed their responses to the researchers rather than the child, framing their remarks as justification rather than dialogue. Across all sessions, only one parent voluntarily sought a child’s input without facilitator prompting. This parent asked, “\textit{Do you want me to see your conversation log with ChatGPT?}” -- a rare attempt to consider the child’s preference in monitoring practices. Most others framed oversight as a non-negotiable responsibility, emphasizing control over collaboration. In effect, children’s presence seldom translated into influence. Parents retained authority and oriented their explanations outward, leaving little room for children’s voices in shaping monitoring practices.

\vspace{2mm}

In summary, parents restrict AI use outside of learning by limiting entertainment time and accompanying children during AI interactions. Despite app-based monitoring, AI is already integrated into daily life, and children access and learn about it through multiple channels. Because of potential negative influences, parents consider child-AI interactions as opportunities to understand their children’s needs and provide guidance when necessary, while avoiding over-monitoring by focusing only on keywords in the content.

\subsection{Parental Perspectives and Pathways in Children's AI Literacy}

This section explores how parents understand children’s AI literacy. The findings reveal that some parents struggle with limited AI knowledge and hold uncertain expectations for their children’s growth in this area. While they emphasize AI mainly for learning purposes and overlook its broader uses and risks, learning contexts offer opportunities for both parents and children to develop AI literacy together.

\subsubsection{Parents' Limited AI Competence and Unclear Expectations}

Eight parents expressed their limited understanding of or limited exposure to AI (H1M, H2M, H5M, H7M, H8M, H10M, H13M, H13D). For example, they could not distinguish between videos generated by AI and those recorded by real people, had difficulty identifying whether apps or devices were AI-integrated, or struggled to verify AI’s answers.

During the study, the mother and child from Household H13 discussed a learning tablet and shared some videos on it. When asked whether these videos were generated by AI or recorded by real people, both H13M and H13K insisted they must have been recorded by real teachers. They based this judgment on seeing a ``teacher’s face'' and hearing ``teacher’s voice,'' unaware of possibilities such as synthetic voices or avatars. 
\begin{quote}
    H13M: \textit{``It’s not animation, it looks like a real teacher.''} \\
    H13K agrees: \textit{``Yes, it’s explained by a real teacher. In some videos, you see the teacher explaining the question. In others, it’s just the teacher’s voice.''} \\
    H13M: \textit{``Right... sometimes it’s like a teacher giving a lecture, standing there as if in a classroom.''}
\end{quote}
Parents' limited AI literacy was also evident in how they described AI's capabilities. H13M believed that \textit{``increasing children’s exposure to AI must be beneficial because some AI functions can surpass humans.''} H8M expected AI to guide the child toward self-discipline and better learning or moral habits through educational AI videos (e.g., tales like Kong Rong Sharing Pears).

Due to their inadequate AI literacy, few parents have concrete, actionable expectations for the children's AI literacy. While they considered self-directness and knowledge base as factors in deciding how much AI exposure their children should have, and some had a rough timeline in mind (Section~\ref{finding_factors}), they were unclear about which competencies children should develop or when these should be acquired. Several parents (H3D, H4D, H8M, H11M, H12D) admitted that they had not thought about it or believed it was too early to consider such questions. In particular, regarding children's future ability to use AI, parents acknowledged the importance of the issue but were unable to answer. As H11M stated,  
\textit{``As for the child’s future ability to use AI and the level they are expected to reach, I have no clear plan or expectation. My attitude is to let things take their natural course.''}

\subsubsection{Perceiving AI Risks Solely Through the Lens of Learning}

In addition to restricting children’s use of AI for learning purposes (Section~\ref{finding_nonlearning}), parents primarily perceived AI as a tool used for learning rather than a general-purpose technology (H2M, H3D, H4D, H5M, H6D, H8M, H11M, H13M). They focused more on how AI could assist children's learning than on fostering children's understanding of its mechanisms, limitations, or appropriate use. 
Consequently, their awareness of AI-related risks was confined to its role in educational support and strongly tied to the context of learning.

Some parents expressed little awareness of risks. They believed that as long as AI could tutor children and clarify knowledge, similar to a private tutor, there was little cause for concern. This reflects some parents' perception of AI as an inherently positive tool for learning, while overlooking potential risks. As H13M explained: \textit{``Concerns? What concerns could there be? If AI can solve problems for the child and explain the concepts clearly, then there’s nothing to worry about. A system like this is basically the same as hiring a private tutor.''}

Other parents recognized potential risks related to children's over-reliance on AI, possible obstacle to the development of critical thinking or creativity, or the risk of receiving incorrect explanations. For instance, H11M emphasized issues of inaccuracy and unreliable guidance due to limited database or imprecise detection of children’s handwriting. 
\begin{quote}
    \textit{``AI is not 100\% accurate. A homework app may return a similar but not identical problem, or even give the wrong explanation if resources are limited or outdated. It can also misread unclear handwriting that parents or teachers would understand. These may be small risks, but they show that accuracy needs to improve. Only as AI becomes more precise will the risks of using it to tutor children go down.''} (H11M)
\end{quote}

When considering AI beyond the educational context, parents were less certain about what risks might emerge or how children would interact with it outside the scope of a ``learning tool.''
Their perspectives ranged from indifference to cautious concern. Some parents regarded understanding AI's mechanisms or limitations as unnecessary, as H13D mentioned:
\begin{quote}
    \textit{``There's no need to understand the principles behind it. As long as the child knows how to use it, that's enough. It’s really that simple. What matters now is being able to use AI, such as DeepSeek.''} (H13D)
\end{quote}
Several parents realized that they had not considered the risks of general AI until asked during the study. They acknowledged the importance of this issue and expressed interest in learning more. Others (H5M, H9M, H11M, H12D) raised concerns about privacy leakage and cyber fraud. They noted that children disclosing excessive personal information to AI might expose sensitive data to misuse or cyberattacks.

\subsubsection{AI Literacy Co-development Through Joint Use of \textcolor{black}{AI} for Learning}

Despite parents' limited AI literacy and their narrow focus on AI as a learning tool, the joint use of AI in learning creates opportunities for both parents and children to improve their AI literacy. Parents are willing to start from learning contexts, using AI tools alongside their children to explore, reflect, and build understanding of AI together.

Five parents (H3D, H4D, H5M, H10M, H12D) described concrete practices of demonstrating AI use cases to their children, guiding children step by step, and reminding them to evaluate and cross-check AI’s responses. For instance, H4D illustrated his own way of using AI with his child and hoped the child could learn from his strategies of iterative questioning, critical evaluation, and reflection. 

\begin{quote}
    \textit{``I show my child how I use AI, like asking for an outline when learning. I remind him answers may be flawed and he must judge with his own knowledge. At work, my first question [to AI] is often clumsy and the answer misses the mark, but I keep refining the prompt and probing from different angles. The response gets sharper. I want him to learn that style of questioning too.''} (H4D)
\end{quote}

Importantly, parents envision this as a continuous, evolving process. As children grow more familiar with AI or as new AI techniques emerge, parents anticipate stepping into the role of co-learners. They imagine learning and using AI together to support schoolwork, as a starting point of improving AI literacy. For example, H5M emphasized encouraging her child to explain reasoning and reflect on AI’s outputs, positioning AI co-use in education as a scaffold for developing critical practices around AI. 

\begin{quote}
    \textit{``I feel I should keep learning so I can help her use AI, that's the stage we're at now. As she grows, if we find useful tools, we’ll learn and use them together to support her learning. She can explain why she wants to use a tool, then we analyze it. If we can’t solve something or want a better answer, we turn to AI. I’ll also have her explain why she agrees or disagrees with AI, building a step-by-step thinking habit.''} (H5M)
\end{quote}

\vspace{2mm}
In summary, some parents demonstrated limited AI literacy where they primarily perceive AI as a tool used for learning and overlook its risks as a general-purpose technology. They also lack clear plans or expectations for cultivating their children's AI literacy. 
However, they expressed a willingness to develop AI literacy alongside their children, beginning within learning contexts.

\section{Discussion}

\subsection{Contextualizing AI Literacy in Families}  

Our findings highlight three distinct ways parents make sense of children’s AI literacy: (1) the absence of a developmental trajectory, (2) the dominance of a pragmatic rather than critical lens, and (3) the emergence of family co-learning as a “third space.” Each highlights different tensions. Parents valued self-regulation and persistence but lacked structured benchmarks to scaffold progress. They emphasized AI’s immediate academic utility over critical reflection on ethics or bias. Yet some families engaged in co-use practices that created more equal dynamics between parents and children, opening space for reciprocal learning. Together, these perspectives suggest that children’s AI literacy in families is currently understood as developmental, pragmatic, and relational -- but in fragmented ways.

\subsubsection{The Absence of a Literacy Trajectory}  
Parents frequently described responsible AI use as dependent on children’s self-discipline, persistence, and independent thinking rather than conceptual knowledge of AI. This emphasis on self-regulation is a positive signal, as it reflects awareness of the metacognitive and motivational foundations of self-directed learning \cite{zimmerman2002self,winne2011self}. Yet it also shifts responsibility away from parents and provides few levers for them to intervene, leaving families uncertain about how to actively foster literacy beyond encouraging independence. In practice, most parents relied on ad hoc indicators such as homework completion or grades to judge competence. This reliance reflects broader struggles in digital literacy assessment \cite{livingstone2011risks}, but it is amplified by AI’s rapid adoption: unlike reading or mathematics, AI literacy lacks shared benchmarks or curricular staging. Without a developmental roadmap, parents cannot reliably scaffold progress and often feel excluded from shaping it.  

The absence of such trajectories has implications beyond the home. Workforce studies consistently highlight AI literacy -- including critical awareness of bias, ethical reasoning, and socio-technical understanding -- as essential for future employability and civic participation \cite{long2020what,ng2021conceptualizing}. If families are left to rely on narrow indicators and self-regulation alone, children may arrive at school-to-work transitions with uneven competencies that privilege functional use over reflective practice. Addressing this gap requires collaboration between researchers, educators, and designers to articulate staged models of AI literacy that align everyday family practices with long-term preparation for workforce and societal demands, much as reading and digital literacies were gradually formalized over time.

\subsubsection{Pragmatic vs. Critical Lens of AI Literacy in Family}  

Most parents in our study viewed AI literacy through a pragmatic lens, treating AI primarily as an academic tool for homework and explanations. This reflects immediate educational pressures and limited parental time, making functional utility the most accessible way to judge value. Such pragmatism resonates with prior work showing that families often prioritize short-term academic outcomes when adopting digital tools \cite{takeuchi2011families,esiyok2025acceptance}.  By contrast, existing frameworks propose a critical lens, emphasizing awareness of bias, ethics, and socio-technical implications \cite{long2020what,ng2021conceptualizing}. These perspectives extend literacy beyond task completion, positioning it as a civic and ethical competency. The gap between the two lenses risks producing children who can use AI effectively but lack the ability to interrogate its outputs or societal consequences.   For HCI, the design challenge is to integrate both lenses. AI systems must remain useful for families’ pragmatic expectations while embedding subtle scaffolds -- such as layered explanations, reflective prompts, or family-facing summaries -- that gradually foster critical engagement. Positioning literacy through dual lenses moves beyond asking which view is correct, toward designing tools that support immediate academic utility while cultivating longer-term reflective capacities.

\subsubsection{Families as Third Spaces for Co-Learning}  

Some families positioned themselves as co-learners with their children, resonating with prior work on families as “third spaces” for technology literacy \cite{druga2022family}. In these settings, co-use was not only about learning how to operate an AI tool but also about negotiating relationships of guidance, trust, and dialogue. Unlike traditional domains where parents serve as primary knowledge holders, children were not yet more fluent with AI than their parents, but many parents anticipated that this shift was inevitable as children grew older and technology became more embedded in their lives. This expectation created a more equal dynamic: parents contributed judgment and contextual understanding, while children brought curiosity and adaptability, setting the stage for reciprocal teaching.  

At the tool level, families collaborated on refining prompts or generating examples for homework, treating AI as a shared resource. At the literacy level, a smaller subset extended these interactions into reflective conversations -- asking together whether an answer was reliable, or discussing when AI should and should not be used. Because such practices depended on parental confidence and children’s willingness to share, many families remained at the tool level. For HCI, this highlights the need to design features that support reciprocity and dialogue -- such as family-friendly reflection cues or shared dashboards -- that allow AI co-use to foster trust and equal participation, helping pragmatic tool use evolve into critical literacy.

\subsection{Design Implications: AI as a Scaffold for Children’s SDL with Parental Support}  
\begin{figure*}[t]
\centering
\begin{tikzpicture}[font=\sffamily, node distance=1cm, scale=0.8, transform shape] 

  \tikzstyle{stage}=[draw, rectangle, rounded corners, minimum width=2.8cm, minimum height=0.9cm, align=center]
  \tikzstyle{labeltop}=[align=center,font=\bfseries]
  \tikzstyle{labelbot}=[align=center,font=\bfseries]

  \node (readiness)   [stage] {Readiness\\ \footnotesize Accurate self-assessment};
  \node (goals)       [stage, right=1cm of readiness] {Goal Setting\\ \footnotesize Clear direction, avoid distraction};
  \node (engagement)  [stage, right=1cm of goals] {Engagement\\ \footnotesize Experimenting with tools};
  \node (evaluation)  [stage, right=1cm of engagement] {Evaluation\\ \footnotesize Reflection on responsible usage};

  \node (gatekeepers) [labeltop, above=0.7cm of readiness] {AI as Gatekeepers \\ \footnotesize Block/delay answers};
  \node (calibrators) [labeltop, above=0.7cm of goals] {AI as Calibrators \\ \footnotesize Unlock modules by literacy};
  \node (facilitators)[labeltop, above=0.7cm of engagement] {AI as Facilitators \\ \footnotesize Summaries, dialogue};
  \node (evaluators)  [labeltop, above=0.7cm of evaluation] {AI as Evaluators \\ \footnotesize Usage logs};

  \node (pmonitor) [labelbot, below=0.7cm of readiness] {Parents as Monitors \\ \footnotesize Setting boundaries, rules};
  \node (pguide)   [labelbot, below=0.7cm of goals] {Parents as Guides \\ \footnotesize Clarity on why and when (not) use};
  \node (ppartner) [labelbot, below=0.7cm of engagement] {Parents as Partners \\ \footnotesize Co-use, conversation};
  \node (pmentor)  [labelbot, below=0.7cm of evaluation] {Parents Handover to Children \\ \footnotesize };

  \draw[->, thick] (gatekeepers.south) -- (readiness.north);
  \draw[->, thick] (calibrators.south) -- (goals.north);
  \draw[->, thick] (facilitators.south) -- (engagement.north);
  \draw[->, thick] (evaluators.south) -- (evaluation.north);

  \draw[->, thick] (pmonitor.north) -- (readiness.south);
  \draw[->, thick] (pguide.north) -- (goals.south);
  \draw[->, thick] (ppartner.north) -- (engagement.south);
  \draw[->, thick] (pmentor.north) -- (evaluation.south);

\end{tikzpicture}

\caption{Compact horizontal mapping: AI roles (top) and Parent roles (bottom) both support four SDL stages (middle).}
\label{fig:sdl-ai-parent}
\end{figure*}
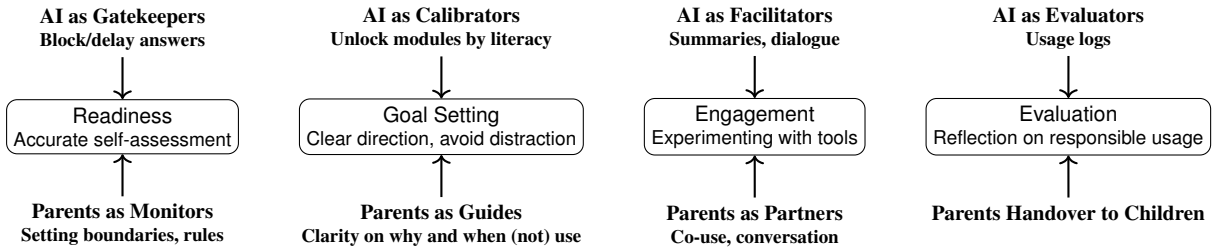

Our findings suggest that parents envision parental control not as a single switch, but as layered functions and roles that scaffold different phases of children’s SDL. By reframing AI as an active partner, these design implications highlight how technology can support readiness, goal setting, engagement, and evaluation in ways that evolve with children’s growth (visualized in Fig. \ref{fig:sdl-ai-parent} ).

\subsubsection{AI as Gatekeepers: Blocking Direct Answers.}
  
Parents emphasized that beyond filtering harmful or inappropriate content, AI systems should be able to block or delay direct answers in order to prevent over-reliance. They worried that if children could always obtain solutions instantly, they would bypass reflection, critical thinking, and effortful problem-solving. Instead, parents envisioned modes where the AI never provides answers outright, or only reveals them after a child’s attempt, emphasizing hints, strategies, or guided feedback. In this model, blocking is not only about safety but also about structuring AI into a learning tool that sustains engagement, builds persistence, and scaffolds SDL. Such features address parents’ concerns that children may otherwise rely too heavily on AI, struggle to self-assess, and fail to initiate the first steps of SDL -- supporting the \textit{readiness to learn} stage.

\subsubsection{AI as Calibrators: Unlocking Features Based on Literacy, Not Age. }
 
Parents stressed that access should expand with children’s competencies rather than chronological age. Because generative AI has become widely available so suddenly, children as young as seven and as old as twelve are encountering it at the same time, making rigid age thresholds especially inadequate. Instead, parents envisioned staged progression: starting with “closed” systems such as learning machines that constrain input, moving toward multiple “small and focused” AI tools (e.g., homework help, writing feedback, exam planning), and eventually unlocking open-ended chatbot features. This modular approach supports purposeful use by encouraging children to identify their needs, consult AI alongside other resources, and practice selecting strategies -- core processes of SDL -- while also making parental guidance and regulation more manageable. Such staged unlocking aligns with educational scaffolding, in which autonomy increases as literacy and responsibility develop, and directly supports the \textit{goal-setting} stage of SDL.

\subsubsection{AI as Facilitators: Fostering Dialogue, Trust, and Co-Learning, Not Policing. }
 
Parents emphasized that effective regulation goes beyond prohibition. They described control features that could foster communication and trust between parent and child, such as activity summaries that prompt conversations or optional co-use of playful learning functions. As children approach adolescence, parents noted that strict prohibitions may backfire, reinforcing resistance rather than responsibility. Instead, they envisioned features that encourage children to explore new functions under parental guidance in ways that are asynchronous and high-level, allowing the system to act as a bridge for dialogue and trust-building. For example, rather than granting parents full access to conversations, the system could provide input summaries or topic lists that help parents understand what children are interested in, creating openings for more meaningful discussions connected to both personal development and AI usage. Parents also highlighted that AI’s rapid pace of change requires \textit{co-learning}: parents need to learn alongside their children to remain able to communicate about new tools and uses, rather than relying solely on control or monitoring. Building on this, when children approach usage boundaries, the system could encourage self-reflection on their patterns and prompt them to engage in dialogue with their parents. These mechanisms, directly mentioned by parents, illustrate how autonomy, oversight, and shared learning can be balanced and directly support the \textit{engagement in the learning process} stage of SDL.

\subsubsection{AI as Evaluators: Making Literacy Visible to Support Independence.  }
Parents emphasized that they often lack the knowledge, access, or time to evaluate their children’s AI literacy, which limits their ability to play an active role in managing AI tool usage. While they can observe outcomes (e.g., grades or completed homework), they have little visibility into how children actually use AI tools, what strategies they rely on, or where they struggle. By contrast, children’s AI interactions generate rich log data, such as how often they request direct answers, whether they use hints before revealing solutions, or how they respond to feedback. Parents noted that such data could provide a more authentic picture of children’s literacy than traditional assessments or standardized questionnaires, which can be easily gamed. Making literacy visible through interaction data allows parents to see “where their child is at” in the SDL process. Over time, these insights can guide a gradual handover of responsibility, enabling children to self-assess, regulate their own strategies, and sustain independent learning with AI -- supporting the \textit{evaluation of learning} stage of SDL.

Together, these perspectives suggest that parental control in children’s AI use should be seen as a developmental spectrum: (1) supporting readiness by blocking risky or answer-giving features so children stay aware of their own learning, (2) guiding goal setting by helping children decide which problems to work on and avoid distraction, (3) supporting the learning process with tools that spark exploration and conversations with parents, and (4) enabling evaluation through both parent-child reflection and data from AI interactions. This spectrum aligns with the four stages of SDL -- readiness, goal setting, engagement, and evaluation -- and points to controls that adapt to children’s growth while keeping parents meaningfully involved.

\section{Limitations}
This study has several limitations. First, while our participants were recruited from different areas within China, the sample size was not large enough to support meaningful comparisons across regions. Families in metropolitan centers, smaller cities, and towns may vary in their access to educational resources, local policies, and cultural practices, but our data do not allow us to draw insights about these differences. Future work should expand participation to larger and more diverse samples to capture these contextual variations. Second, our contributions primarily reflect parents’ perspectives. Although children were present in the sessions and their voices were partially represented through literacy questions and discussions, they were not direct participants in envisioning the future of AI use. Parents in our study were also less comfortable with researchers engaging children alone about AI usage, suggesting that new methodological strategies are needed to foreground children’s perspectives in subsequent work. Third, our study design emphasized the family as the unit of analysis, privileging parental mediation practices. This approach risks underrepresenting the influence of other stakeholders, such as teachers or peers, whose practices also shape how children develop AI literacy. Future research should triangulate across these perspectives to build a more holistic understanding.

\section{Conclusion}

Generative AI introduces both opportunities and risks for children’s self-directed learning (SDL). Through focus groups with parents and children (ages 7--13) across 13 households, we found that parents conceptualize children’s AI literacy as progressing through screen time, self-directness, and knowledge growth, yet primarily confine AI to learning contexts. Their monitoring practices, such as regulating apps or contexts, often conflicted with AI’s seamless infrastructural presence in daily life. Parents also reported limited competence and uncertain expectations, treating co-use as opportunities for co-learning. Critically, parents tended to view AI literacy through a pragmatic lens of homework help, overlooking the critical dimensions of ethics, bias, and broader socio-technical risks. By emphasizing self-discipline and independence, they shifted responsibility onto children while lacking developmental roadmaps to scaffold progress. These findings situate children’s AI literacy within families and highlight design needs for responsible AI that not only supports phased SDL but also bridges pragmatic expectations with critical literacies and long-term developmental goals.

\bibliographystyle{style.bst}  
\bibliography{text.bib}  

\newpage
\appendix

\section{AI for SDL Questions} \label{tenquestion}
\vspace{-1em}

\begin{table*}[h]
\centering
\caption{AI Literacy and Self-Directed Learning (SDL) Assessment (Middle School)}
\label{tab:ai_sdl_questions}
\begin{tabularx}{\textwidth}{@{}p{0.3cm} p{3.4cm} p{6cm} p{4cm}@{}}
\toprule
\textbf{ID} & \textbf{Question} & \textbf{Scoring (2 / 1 / 0)} & \textbf{Framework Aspect \& References} \\
\midrule
\multicolumn{4}{@{}l}{\textbf{Understanding of AI (AI Literacy)}}\\
\midrule
Q1 & What is Artificial Intelligence (AI)? &
2 = Explains AI as a computer system/program that learns from data to make predictions or create content; 
1 = Vague answer like “a smart computer/robot”; 
0 = Wrong or no answer. &
\textbf{Conceptual knowledge of AI} -- what AI is; definitional clarity \cite{long2020what,touretzky2019ai4k12,unesco2021aicompetency}. \\
\addlinespace[0.4em]
Q2 & Where do you see AI in daily life? Give two examples. &
2 = Gives two valid examples (e.g., voice assistant, recommendation system, ChatGPT); 
1 = Only one valid example; 
0 = No valid example. &
\textbf{AI awareness \& contexts of use} -- everyday encounters with AI \cite{touretzky2019ai4k12,long2020what}. \\
\addlinespace[0.4em]
Q3 & What is AI good at, and what is it not good at? &
2 = Mentions at least one strength and one limitation; 
1 = Only a strength or only a limitation; 
0 = No response. &
\textbf{Capabilities \& limitations} -- strengths, failure modes, bias/accuracy concerns \cite{long2020what,unesco2021aicompetency}. \\
\addlinespace[0.4em]
Q4 & How does generative AI (like ChatGPT) learn to answer questions? &
2 = Learns patterns from large amounts of text/data but doesn’t truly “understand”; 
1 = Says “it learns from people” but vague; 
0 = Wrong (e.g., “it thinks like a human”). &
\textbf{How AI works (simplified)} -- data, patterns, training \cite{touretzky2019ai4k12,long2020what}. \\
\addlinespace[0.4em]
Q5 & What could happen if you share too much personal information with AI? &
2 = Mentions specific risks (data leaks, scams, safety issues); 
1 = Says “it’s unsafe” without explanation; 
0 = No risks mentioned. &
\textbf{Ethics, privacy, \& safety} -- responsible use \cite{unesco2021aicompetency,long2020what}. \\
\midrule
\multicolumn{4}{@{}l}{\textbf{AI and Self-Directed Learning (SDL)}}\\
\midrule
Q6 & If AI gives you an answer, how do you check if it’s correct? &
2 = Cross-check with other sources (teacher, books, reliable websites) or reason it out; 
1 = Just ask AI again or trust it directly; 
0 = No way to check. &
\textbf{Motivation / Readiness} willingness to take responsibility for accuracy \cite{knowles1975self,garrison1997self}. \\
\addlinespace[0.4em]
Q7 & If you use AI to prepare for exams, how would you use it? &
2 = Have AI make practice questions, explain step by step, or plan review sessions; 
1 = Just ask AI for answers; 
0 = No method. &
\textbf{Goal-setting} planning, structuring, and directing learning activities \cite{knowles1975self,garrison1997self}. \\
\addlinespace[0.4em]
Q8 & If you don’t understand AI’s answer, what do you do next? &
2 = Rephrase, keep asking, or look it up elsewhere; 
1 = Ignore it or ask the teacher; 
0 = No response. &
\textbf{Self-monitoring / Engagement} adapting strategies to maintain learning progress \cite{garrison1997self}. \\
\addlinespace[0.4em]
Q9 & How do you write a good question (prompt) so AI gives a more useful answer? &
2 = Be clear, specific, and give details; 
1 = Just ask randomly; 
0 = No response. &
\textbf{Self-management / Engagement} selecting and applying effective learning methods \cite{knowles1975self,garrison1997self}. \\
\addlinespace[0.4em]
Q10 & How do you know if you’re relying too much on AI instead of really learning? &
2 = Mentions signals (always copying, not understanding, not remembering); 
1 = Vague answer (e.g., “using it too much”); 
0 = No response. &
\textbf{Self-monitoring / Evaluation} reflecting on independence and quality of learning outcomes \cite{garrison1997self}. \\

\bottomrule
\end{tabularx}
\end{table*}

\end{document}